
\relax
\tolerance=500
\magnification=\magstep1
\hoffset=0.5cm
\voffset=0.5cm
\baselineskip=18pt plus .1pt
\def\neq{\hbox{$\,$=\kern-6.5pt /$\,\,$}}
\baselineskip=23 pt plus 1 pt

\centerline{\bf STATIONARY AXISYMMETRIC FIELDS AS TWO-DIMENSIONAL
GEODESICS}
\vskip 1cm
\centerline{Dar\'\i o N\'u\~nez$^{\dagger }$ and Hernando Quevedo}
\vskip.3cm
\centerline{Instituto de Ciencias Nucleares}
\centerline{Universidad Nacional Aut\'onoma de M\'exico}
\centerline{A. P. 70--543}
\centerline{04510 M\'exico, D. F., MEXICO}
\centerline{$^{\dagger}$ Present Address: Theoretical Physics
Institute, University of Alberta, }
\centerline{Edmonton, Alberta T6G 2J1, Canada.}
\vskip.3cm
\vskip 2cm
\centerline{\bf ABSTRACT}

Einstein's equations for stationary axisymmetric fields are
reformulated
as the equations for affine geodesics in a two--dimensional space.
The affine collineations of this space are investigated and used
to relate explicit solutions of Einstein's equations with different
physical
properties. Particularly, the solutions describing the exterior
fields of a
dyon and a slowly rotating body are discussed.

\vfill\eject\noindent
{\bf 1. Introduction}

To simplify the structure of Einstein's equations, it is usual to
postulate
the existence of one or more Killing vector fields in the spacetime
under
consideration or, in less technical terms, the independence of
certain
coordinates. In a more general sense, the omission of the coordinates
can
be regarded as a special case of the Kaluza--Klein approach. Indeed,
to
investigate solutions with two Killing vectors in a systematic
fashion,
we can consider a Kaluza--Klein type reduction of Einstein's theory
to two
dimensions [1]. The dimensional reduction just amounts to dropping,
for all
the fields in the spacetime, the dependence on the coordinates that
can be
associated with the Killing vectors.

In this work, we are concerned with a different type of dimensional
reduction in which the number of fields -- in our case, the metric
coefficients -- is reduced to the minimum necessary for describing
the spacetime. This reduction occurs at the level of the
Einstein--Hilbert
Lagrangian and consists in dropping the terms that can be represented
as total divergences, and rearranging the non--ignorable terms so
that the
Lagrangian becomes two--dimensional. The spacetime coordinates are
absorbed into certain differential operators that act on the
remaining
metric coefficients, i.e., the metric coefficients become the
coordinates of

the reduced space.  This idea is in the spirit of the construction of
the

superspace studied by de Witt and others [2], where each point is a

space. In our case, each geodesic defines a solution to the

Einstein equations.

Neugebauer and Kramer [3] introduced the abstract potential space

determined by the Einstein--Hilbert Lagrangian of Einstein--Maxwell
fields,
and investigated the field equations which are derivable from a
minimal
surface problem in the potential space. In a recent work [4], we
showed that
canonical transformations can be used to reduce the dimensionality of
the

potential space, and Einstein's equations coupled to any matter field
are

equivalent to the geodesic equations in a two--dimensional space.

As it is known,
there exists a great deal of hypersymmetry in bidimensional physics,
that is,
supposedly unrelated problems happen to be the very same one, once a
type of conformal transformation is given [5]. In our geodesic
problem,

this is equivalent to relate, or generate, new solutions to the
affine

geodesic motion, which, again refrased in the Einstein equations,
translates
into the generation of solutions.

We will focus attention on stationary axisymmetric solutions to the
Einstein
equations. In Section 2, the  corresponding field equations are
derived from
a two--dimensional metric Lagrangian and it is shown that they may be
interpreted as the equations for an affine geodesic. We then
investigate the
equation for affine collineations and present the general solution
for the
special case of a symmetry vector that depends on the coordinates
only.
Section 3 contains a solution generated by applying three different
transformations on the Chazy--Curzon metric. We study the properties
of
this solution and show that it may be interpreted as describing the
exterior field of a gravitational dyon. Section 4 is devoted to the
derivation and study of a solution which contains the parameters
necessary
to describe the field of a slowly rotating mass.

\vskip.3cm\noindent
{\bf 2. Field equations and affine collineations}

Consider the general stationary axisymmetric line element in Weyl
canonical coordinates
$$ds^2 = e^{2\psi}(dt-\omega d\phi)^2 -e^{-2\psi}
[e^{2\gamma}(d\rho^2 + dz^2 )+ \rho^2 d\phi^2]\ , \eqno(1)$$
where $\psi,\ \omega$, and $\gamma$ are functions of $\rho$ and $z$
only.
If $\omega =$const., Eq.(1) leads to the special case of static
axisymmetric fields.
The calculation of the corresponding scalar curvature leads to the
Lagrangian
$$L = {e^{4\psi}\over 2\rho} (\omega_\rho^2 + \omega_z^2)
 + 2\rho (\psi_{\rho\rho} + \psi_{zz} - \gamma_{\rho\rho} -
\gamma_{zz}
- \psi_\rho^2 - \psi_z^2) + 2\psi_\rho\ , \eqno(2)$$
which generates the usual Einstein equations. We proceed to give the
main
steps to construct from Eq.(2) another Lagrangian which generates the
equation for affine geodesics.

Introducing  the differential operator $ D  = (\partial_\rho, \
\partial_z)$,
Eq.~(2) can be written as
$$L = 2 D \rho  D \gamma + {e^{4\psi}\over 2\rho}( D  \omega)^2
-2\rho ( D  \psi)^2 \ . \eqno(3)$$
In obtaining Eq.(3), we made the substitution $\rho  D^2 B =
 D (\rho D  B) -  D  \rho  D  B$ and neglected the total
divergence terms. The Lagrangian (3) may now be interpreted as
describing a
kinematic system defined in the three-dimensional ``space of metric

coefficients" with generalized ``coordinates" $\psi, \ \omega$, and
$\gamma$ that depend on $\rho$ and $z$. Consequently, $\rho$ and $z$
may be
considered as quantities used to parametrize the coordinates.
Equation (3)

shows that in this special case the Lagrangian explicitly depends on
the

parameter $\rho$. Now, since $\gamma$ and $\omega$ are cyclic
coordinates of

the Lagrangian (3), it is convenient to use the Routhian $R$ obtained
from
the Lagrangian $L$ by means of Legendre transformation acting on the
cyclic

coordinates only, i.e.,

$$R = {\partial L\over \partial (D\gamma)} D\gamma +
      {\partial L\over \partial (D\omega)} D\omega - L =
{1\over 2} \rho e^{-4\psi} \Pi_\omega^2 + 2\rho (D\psi)^2\ .
\eqno(4)$$
Here $\Pi_\omega $ is the canonically
conjugate ``momentum" associated with the generalized coordinate
$\omega$.

Note that the conjugate momentum $\Pi_\gamma$ (as well as $\gamma$)
does not enter the Routhian (4)
at all. As a consecuence, it can be shown that the metric function
$\gamma$
is determined by two first
order partial differential equations that can be integrated by
quadratures
once $\psi$ and $\omega$ are known [6].

It follows from Eq.(4) that $\Pi_\omega$ is a ``constant of motion"
 (i.e., $D\Pi_\omega =0$) in space of metric coefficients. Using this
fact, we
can define an additional differential operator
$\widetilde D = (-\partial_z, \ \partial_\rho)$
such that $D \widetilde D \equiv 0$. Introducing a function $\Omega$
by means
of the relationship
$$\Pi_\omega = \rho^{-1} e^{4\psi} D\omega =\widetilde D \Omega \
,\eqno(5)$$
the Routhian (4) becomes
$$ R = {1\over 2} \rho f^{-2} [ (D\Omega)^2 + (Df)^2] \ , \eqno(6)$$
where $f=$exp$(2\psi)$. In this way we have obtained a new
Lagrangian,
Eq.~(6), which corresponds to  the squared line
element of an abstract  space described by $f$ and $\Omega$ and
endowed
with a two--dimensional metric tensor conformal to the Euclidean one,
i.e.,
$$g_{ab} ={1\over 2} \rho f^{-2}\left( \matrix{ 1 & 0 \cr
                                             0 & 1 \cr }
\right).\eqno(7)$$
The metric functions $f$ and $\Omega$ plays the
role of coordinates, and the operator $D$ may be interpreted as the
derivative  with respect to the affine parameter(s) used to
parametrize
the coordinates $f$ and $\Omega$. Consequently, the spacetime
coordinates
turn out to be affine parameters on the space of metric coefficients.

Therefore, Eq.(6)
determines a metric Lagrangian that explicitly depends on the affine
parameter $\rho$.

The geometric non--vanishing quantities associated with the metric
(7) are,

up to symmetries,
$$\eqalign{
\Gamma^1_{\ 11} =- \Gamma^1_{\ 22} = \Gamma^2_{\ 12} &=-{1\over f},\
\cr
R^1_{212} &={1\over {f^2}} \, , \cr
R_{11}=R_{22} &= {1\over {f^2}} \, , \cr}
\eqno(8)
$$
and the scalar curvature is $4/\rho$.

The Euler--Lagrange motion equations obtained from the Routhian (6)
are
$$
\eqalign{ D^2 f -f^{-1}(Df^2 - D\Omega^2) + \rho^{-1}D\rho Df &=0,
\cr
 D^2\Omega - 2f^{-1}Df D\Omega + \rho^{-1}D\rho D\Omega &= 0, \cr}
\eqno(9)$$
which are the same principal equations which
follow  from $R_{\mu \nu}=0$. For completeness, we mention that
taking
$E = f + i \Omega $, the Routhian (6) can be rewritten as
$$R ={2  \rho \over (E + E^*)^2 }\ DE \ DE^* \ , \eqno(10)$$
where an asterisk represents complex conjugation. The variation of
Eq.~(10)
 with respect to $E$ or $E^*$ leads to the Ernst equation [7]
$$ ({\rm Re} \, E) \Delta E = (DE)^2 , \qquad {\rm with} \qquad
\Delta E = D^2 E + \rho^{-1} D\rho D E \ . \eqno(11)$$

Since the starting Routhian (6) depends
explicitly on the parameter $\rho$, Eqs.(9) coincide with the
equations
for an {\it affine} geodesic

$$D^2X^a + \Gamma^a_{\ bc} DX^b DX^c = \lambda(\rho) DX^a \ . $$
An affine geodesic
is therefore a two--dimensional curve $ X^a= (f, \ \Omega)$ with
tangent
vector $DX^a$ satisfying Eqs.(9) for the real function $\lambda(\rho)
=
-1/\rho$. The existence and uniqueness of solutions of Eqs.(9) follow
from theorems on systems of differential equations. Let $X^a_0$ be a
point
of the space generated by $g_{ab}$ and $DX^a_0$ the value of the
tangent

vector at
this point. Then, there exists a unique, up to a change of the
parameter,
maximal affine geodesic $X^a$ such that $X^a(0)=X^a_0$ and $DX^a(0) =
DX^a_0$
(see, for instance, Ref. [8]). This result represents an alternative
proof
of the fact that a stationary axisymmetric solution is uniquely
determined
by its values on the axis of symmetry [9].

Let $X^a_1$ and $X^a_2$ represent two different affine geodesics. Our
goal
is to find transformations that relate $X^a_1$ with $X^a_2$ and may
be used
to generate new solutions from known ones. The existence of this type
of
transformations cannot be assumed {\it a priori} and it depends on
the
symmetry properties of the underlying equations as well as on the
explicit
form of $X^a_1$ and $X^a_2$. In the four--dimensional spacetime,
transformations generating new solutions have been extensively
studied and
applied to diverse problems [10]; here, we first carry out the
dimensional
reduction and then reduce the problem to that of affine geodesics. To
begin
with the study of the transformations relating two different
solutions of
Eqs.(9), we consider the simplest case of a linear infinitesimal
transformation.

For the general affine geodesic equation,
$$D^2 X^a + \Gamma^a_{\ b c}DX^b DX^c + g^{ab}D(g_{bc})DX^c= 0 \ ,
\eqno(12)$$
an infinitesimal transformation
$$X^a \rightarrow {X^\prime}^a = X^a + \epsilon\, \eta^a\ ,
\eqno(13)$$
is said to be a symmetry transformation, that is, maps solutions into
solutions, to order $\epsilon$, if the symmetry vector $\eta^a$
satisfies
the condition
$$\bar D^2 \eta^a + R^a_{\ b c d}\, DX^b\, DX^c\, \eta^d
-(\Gamma^a_{\ b c}
)_{,s} \, DX^b \, \eta^c = 0\ , \eqno(14)$$
where $\bar D$ is the total derivative operator on shell
$$\bar D={\partial \over {\partial s}} + DX^a{\partial \over
{\partial X^a}} -
[\Gamma^a_{\ b c}DX^b DX^c + g^{ab}D(g_{bc})DX^c]{\partial \over
{\partial DX^a}}\ , \eqno(15)$$
and $s$ is a parameter along the geodesics. Moreover,  $ R^a_{\ b c
d}$ is the
Riemann tensor of the space of metric coefficients and period stands
for

partial derivative.
If the metric Lagrangian is independent of the parameter $s$,
Eq.(14) reduces to the equation of affine collineations ${\cal L}
\Gamma^a_{\ bc} = 0$, where ${\cal L}$ is the Lie derivative along a
curve
with tangent vector $\eta^a$, which is equivalent to the equation of
the
geodesic deviation for the connecting vector $\eta^a$. Hence the
geometrical basis of our approach becomes plausible. A family of
solutions
of Einstein's equations is equivalent to a  congruence of geodesics
in the
space of metric coefficients. If $\eta^a$ is a vector connecting two

neighboring geodesics
at a given point,  then the condition for $\eta^a$ to remain a
connecting
vector at any other point of the space of metric coefficients, at
which the

geodesics are
well--defined, is that it must satisfy the equation of geodesic
deviation.

If the symmetry vector $\eta^a$ is just a function of the parameter
$s$ and

the coordinates, then the symmetry equation (14) can be rewritten as
$$\eta^a_{,ss} + 2 (\eta^a_{,s})_{;b} DX^b +
(\eta^a_{; b c} + R^a_{\ b c d}\,\eta^d ) DX^b\, DX^c\, = 0\ ,
\eqno(16)$$
where a semicolon represents the covariant derivative associated with
the
metric $g_{ab}$ given in Eq.(7). Notice that in the case that
$\eta^a$ is just
a function of the coordinates, even for a metric depending on the
non--affine
parameter, the symmetry equation reduces to that of affine
collineations.
Consider this last case, $\eta^a = \eta^a ( X^b )$. Introducing the

metric (7) into the symmetry equation (16), we get
$$\eqalign{
D^2\eta^1 -2 f^{-1}(Df D\eta^1 - D\Omega D\eta^2) + \eta^1 f^{-2}
[(Df)^2 - (D\Omega)^2] + \rho^{-1} D\rho D\eta^1 &= 0\ , \cr
D^2 \eta^2 - 2 f^{-1} (Df D\eta^2 + D\Omega D\eta^1) + 2 \eta^1
f^{-2}
Df D\Omega + \rho^{-1}D\rho D\eta^2 &= 0 \  . \cr}
\eqno(17)$$
A detailed investigation of Eq.(17) shows that it possesses three
independent
solutions:
$$ \eta^a_1 = ( 0, 1 ) \ , \eqno(18)$$
$$ \eta^a_2 = ( f, \Omega ) \ , \eqno(19)$$
$$ \eta^a_3 = ( f\, \Omega, {{\Omega^2 - f^2}\over 2} ) \ .
\eqno(20)$$
Moreover, it can be shown that there are no affine
eigencollineations,
that is, the solutions (18--20) coincide with the Killing vectors of
the metric (7). To find more general symmetry vectors of the
potential
space, it is necessary to consider the most general {\it ansatz}
$\eta^a = \eta^a(s,X^b, DX^b)$. In this work, however, we want to
focus
attention on the symmetry vectors (18--20) and to show that even
these
simple vectors can be used to connect classes of solutions with
different
physical properties.

We will now consider the type of solutions which can be generated
by means of the vectors (18--20). Let $\epsilon_1$, $\epsilon_2$, and
$\epsilon_3$ be the parameters introduced by the symmetry vectors
$\eta^a_1$, $\eta^a_2$, and $\eta^a_3$, respectively,  according to
Eq.(13).
Acting on a seed solution ($f, \Omega$), the vector $\eta^a_1$ leads
to
the new affine geodesic $f' = f$ and $\Omega' =\Omega + \epsilon_1$.
According to Eq.(5), this
is equivalent to adding a constant $\omega_0$ to the metric function
$\omega$.
Obviously, this symmetry transformation is trivial since a coordinate
transformation of the form $t' = t - \omega_0 \phi$ in the line
element (1)
absorbes the new term. Physically, this is equivalent to the
introduction
of a rotating frame for the line element (1). Similarly, it is
possible
to show that the parameter $\epsilon_2$ associated with the symmetry
vector
$\eta^a_2$ can be absorbed by means of a rescaling of coordinates.
The only
non--trivial symmetry vector is $\eta^a_3$ and it can be used to
generate
new solutions of the form
$$ f' = f(1+\epsilon_3 \Omega)\ , \qquad \Omega' = \Omega
  + {\epsilon_3\over 2}(\Omega^2 - f^2) \ . \eqno(21)$$

Although, when acting alone, the symmetry vectors $\eta^a_1$ and
$\eta^a_2$
are trivial, we will see below that they are helpful when used
together with
$\eta^a_3$ to generate non--trivial solutions.

 Note, moreover, that the
corresponding parameters $\epsilon_1$ and $\epsilon_2$ can take any
real
value  because they do not enter the symmetry equations at all. That
is,

putting the infinitesimal transformation (13) with $\eta^a_1$ and
$\eta^a_2$,
one sees that the resulting equation is identically satisfied
regardless
of the values of the parameters $\epsilon_1$ and $\epsilon_2$,
respectively.

\vskip.3cm\noindent
{\bf 3. Exterior field of a gravitational dyon}

The interest in monopole structures has rapidly increased during the
past
few years due to their discovery in generalizations of the standard
model
of particle physics. Magnetic monopoles were first introduced by
Dirac [11]
in electrodynamics to symmetrize Maxwell's equation in a direct way.
Certainly, the most important consequence of the existence of
magnetic
monopoles is the quantization of electric charge. Most grand unified
theories possess t'Hooft--Polyakov monopoles [12]. In general
relativity
there exist two different sorts of monopole structures: a
magnetically
charged black hole and a gravitational dyon. In fact, the magnetic
black
hole is the magnetic counterpart of the electrically charged black
hole

described
by the Reissner--Nordstrom metric, and is related to it by a duality
rotation. A magnetic black hole can also be interpreted as a magnetic
monopole with mass greater than a determined critical value [13].

A gravitational dyon is a hypothetical object the existence of which
follows
from the {\it relativistic} character of gravitation. In Newtonian
theory,
the only source of gravitation is the mass. In contrast, general
relativity
predicts that mass as well as rotation are stationary sources of
gravitational interaction. This leads to the well--known analogy
between
relativistic gravity and electromagnetism. The gravitational field
generated
by a distribution of mass turns out to be analogous to the electric
field,
and the field of an angular momentum current presents characteristics
similar
to those of a pure magnetic field. For this reason, the field
generated by an
angular momentum current is called ``gravitomagnetic" field. For this
analogy to be complete, it is necessary to require the existence of a
``gravitomagnetic monopole" as the counterpart of the magnetic Dirac
monopole
of electrodynamics. A gravitational dyon is thus a mass endowed with
a
gravitomagnetic monopole. In this section, we will investigate
solutions
that can be generated from a static seed metric by means of a
combination
of symmetry transformations, and may be used to describe the exterior
field
of a gravitational dyon.

To give a correct interpretation of the solutions presented here, we
will
use a coordinate--invariant method based upon the investigation of
the
relativistic multipole moments for asymptotically flat solutions,
according
to the definition proposed by Geroch and Hansen [14]. We now proceed

to derive the solution for a gravitational dyon. If we consider a
static
asymptotically flat solution $(f, \Omega=0)$ as seed metric and apply
to it
the symmetry transformation associated with the vector $\eta^a_3$, we
obtain a stationary solution with $f' =f$ and $\Omega' = -\epsilon_3
f^2/2$.
It can be shown that for any given asymptotically flat $f$ the new
solution
does not satisfy the condition of asymptotic flatness \'a la
Geroch--Hansen
[15]. Consequently, it is not possible to covariantly interpret the
solutions
generated by this type of transformation. To avoid this difficulty,
we use
a combination of three different symmetry transformations (18--20).
To the
seed static solution $f$ we first apply the symmetry vector
$\eta^a_1$ with
parameter $\epsilon_1$. The resulting solution is then used as seed
solution
for a transformation with the vector $\eta^a_2$ and parameter
$\epsilon_2$,
and, finally, we apply the symmetry vector $\eta^a_3$. The new
solution
can be written as
$$f' = (1+\epsilon_2) f [ 1 + \epsilon_1\epsilon_3 (1+\epsilon_2) ]\
,
\eqno(22)$$
and
$$\Omega' = (1+\epsilon_2)\left[ \epsilon_1 -{\epsilon_3\over
2}(1+\epsilon_2)
(f^2 - \epsilon_1^2)\right]\ . \eqno(23) $$
It is now necessary to choose the parameters introduced by the
symmetry
transformations such that the new solution becomes asymptotically
flat. This
condition leads to the relationships
$$ \epsilon_1^2 = -{\epsilon_2\over 2+\epsilon_2 }\ ,
\quad {\rm and} \quad
\epsilon_3 = -{\epsilon_2\over \epsilon_1 (1+\epsilon_2)^2 }\ ,
\eqno(24)$$
where $\epsilon_2$ is a negative constant defined in the interval
$\epsilon_2
\in (-2, 0)\backslash \{-1\}$. As we mentioned at the end of section
2, the
parameters $\epsilon_1$ and $\epsilon_2$ do not need to be
infinitesimally
small. Consequently, they can be chosen such that Eq.(24) is
satisfied and
$\epsilon_3$ becomes infinitesimally small as required by the
transformation
law (13). In fact, even for very large values of $\epsilon_1$,
$\epsilon_3$
remains infinitesimal and $\epsilon_2$ remains in its domain of
definition.

To analyze a concrete solution, we have to specify the asymptotically
flat
seed metric. Consider the Chazy--Curzon metric [16]
$$ f = \exp(-2m/r)\ , \qquad r^2 = \rho^2 + z^2 \ , \eqno(25) $$
where $m$ is a positive constant. The new solution is then given by
Eqs.(22),
(23) and (25). Choosing the new parameters according to Eq.(24), we
calculate
the corresponding Geroch--Hansen multipole moments and obtain
$$ M_0 = m \ , \qquad J_0 = - m \epsilon_3 \ . \eqno(26)$$
There are higher mass multipole moments $M_n$ which corresponds to
the
axisymmetric mass distribution of the source, and higher moments for
the
angular momentum current $J_n$ which, however, can be neglected since
they
are proportional to $\epsilon_3^2$. Equation (26) shows that this
solution
represents the gravitational field of a body with mass $m$ and
gravitomagnetic
monopole $-m\epsilon_3$. Hence, the new parameter $\epsilon_3$ may be
interpreted as the specific ``gravitomagnetic" mass which may be
positive as
well as negative. The total ``gravitoelectric" mass of the seed
solution has
not been affected by the action of symmetry transformations. For the
sake of completeness, we present the metric functions of the new
solution:
$$ f' = \exp(-2m/r)\ ,  \quad \omega' = -2m\epsilon_3(1+\epsilon_2)^2
z/r\ ,
\quad \gamma' = -m^2\rho^2/r^4 \ . \eqno(27)$$

Finally, we would like to mention that using the Schwarzschild metric
as starting solution, it is possible to generate the linearized
Taub--NUT
(Newman--Unti--Tamburino) solution which is also a candidate for
describing
the exterior field of a gravitational dyon. In general, it should be
possible
to find other solutions which, being different from the Taub--NUT
metric or
the one presented here, present similar properties and hence might be
used
to describe a dyon. They all could differ only in the set of
multipole
moments higher than the monopole one; that is, there may exist
different
distributions of mass possessing the same gravitomagnetic monopole
structure.

\vskip.3cm\noindent
{\bf 4. Field of a slowly rotating mass}

For the study of the gravitational field of astrophysical bodies like
stars and planets, it is necessary to investigate solutions which
possess
a set of mass multipole moments as well as a set of gravitomagnetic
moments
representing the rotation of the source. In contrast to the solution
presented in the last section, a solution with realistic rotational
properties may have only gravitomagnetic multipoles higher than or
equal to
the dipole one. In this section we derive a solution wich satisfies
this
condition.

Consider any stationary seed solution $(f, \Omega)$ satisfying the
conditions
of asymptotic flatness. As we have done
in section 3, we apply three consecutive symmetry transformations
according
to Eqs.(13) and (18--20). The new solution is then given by
$$f' = (1+\epsilon_2) f [ 1 +\epsilon_3 (1+\epsilon_2) (\Omega +
\epsilon_1) ]
\ ,
\eqno(28)$$
$$\Omega' = (1+\epsilon_2)\left[ \Omega + \epsilon_1 -
{\epsilon_3\over
 2}(1+\epsilon_2)
(f^2 - \epsilon_1^2-2\epsilon_1\Omega - \Omega^2)\right]\ . \eqno(29)
$$
In general, this new solution is not asymptotically flat. However, if
we
demand that the parameters $\epsilon_1$ and $\epsilon_2$ satisfy the
relationships (24), asymptotic flatness is conserved and the
resulting
solution can be written as
$$ f' = f[ 1 + \epsilon_3 (1+\epsilon_2)^2 \Omega ]\ , \eqno(30)$$
$$\Omega' = \Omega + {\epsilon_3 \over 2} (1 + \epsilon_2)^2 ( 1 +
\Omega^2
-f^2) \ . \eqno(31)$$

The calculation of new solutions does not present any difficulties.
We will
present here only one solution which illustrates our approach and can
easily
be interpreted. Consider the seed solution [17]
$$ f = {x^2  -1 + \alpha_1^2 (y^2-1)\over (x+1)^2 + \alpha_1^2
(y-1)^2 }\ ,
\qquad
\Omega = {2\alpha_1 (x+y)\over  (x+1)^2 + \alpha_1^2 (y-1)^2 }\ ,
\eqno(32)$$
with
$$ x = {1\over 2m} (r_+ + r_-)\ , \qquad y={1\over 2m}(r_+ - r_-)\ ,
\qquad r_\pm^2 = \rho^2 + (z\pm m)^2 \ , $$
where $m$ and $\alpha_1$ are constants. To illustrate the effect of
symmetry
transformations, we first analyze the seed solution (32). An
investigation
of the corresponding multipoles show that there are gravitoelectric
as well
as gravitomagnetic monopole and dipole moments. Due to the presence
of the
gravitomagnetic monopole and gravitoelectric dipole, this solution
cannot
be considered as a candidate for the description of the gravitational
field
of any astrophysical object. Hence solution (32) is of no interest
from
a physical point of view. However, if we apply three different
symmetry
transformations to solution (32), its physical meaning can totally be
changed.
In fact, putting Eq.(32) into Eqs.(30) and (31), and calculating the
relativistic multipole moments of the resulting solution, we see that
all undesirable multipole moments vanish if $\alpha_1$ is assumed to
take
the value
$$\alpha_1 = -\epsilon_3(1+\epsilon_2)^2\ . \eqno(33)$$
Then, the only nonvanishing multipoles are
$$M_0 = m \ , \qquad {\rm and} \qquad J_1 = \epsilon_3 (1+
\epsilon_2)^2 m
\ . \eqno(34)$$
The last equation shows that the total mass of the body is given by
$m$ and
that only the gravitomagnetic dipole moment survives in accordance
with the
dipole character of rotation. The angular momentum per unit mass is
given
by $\epsilon_3(1+\epsilon_2)^2$ and can be positive as well as
negative,
corresponding to the two possible directions of rotation of the
source
with respect to the symmetry axis.  Consequently, the new solution
may be
interpreted as describing the exterior field of a slowly rotating
mass.
Using Eqs.(30)--(33) and (5), the calculation of the metric
components leads
to
$$ f' = {x-1\over x+ 1} \ , \quad \omega' =
2m\epsilon_3(1+\epsilon_2)^2
{ 1 - y^2 \over x-1} \ , \quad \gamma' = {1\over 2} \ln {x^2-1\over
x^2-y~2}
\  . \eqno(35)$$
This is equivalent to the Lense--Thirring metric [18], the physical
meaning
of which  has been investigated by using other approaches and
coincides with
that we have obtained above by just analyzing the corresponding
multipole
moments.

\vfill
\eject
\noindent
{\bf 5. Conclusions}

We have presented a different way to view to Einstein's

equations, mainly as geodesic motions in a space where the metric

coefficients of the spacetime play the role of coordinates.  The

approach presented here for the axisymmetric stationary case can be

generalized to any spacetime and even to the non-vacuum cases. We

have reasons to believe that a dimensional reduction, via canonical

transformations of the Hamiltonian, can always be made in this space

of metric coefficients, and field potentials, such that the dynamical

problem reduces to study the geodesic motion in a two dimensional

manifold.

As an application of this point of view we studied the symmetries of

the geodesic motion for the space associated with axisymmetric

stationary gravitational fields, and were able to generate some
solutions,

whose

implications are currently under study. Nevertheless, we want to

stress the fact that the idea presented here is not only a method for

generating solutions but more than that, a different point of view to

work with Einstein's equations.

\vskip 1cm
\centerline{\bf References}
\parindent=.8cm
\vskip .6cm
\item{[1]} B. Julia, in {\it Superspace and supergravity}, edited by
S.W. Hawking and M. Rocek (Cambridge University Press, 1980).

\item{[2]} B. S. de Witt, Phys. Rev. {\bf 160}, 1113 (1967).

\item{[3]} G. Neugebauer and D. Kramer, Ann. Phys. (Leipzig) {\bf 24}
62
(1969).

\item{[4]} D. N\'u\~nez and H. Quevedo, to be published in
``Proceedings of

the Conference SILARG VIII'' (Campinas, Brazil, 1993).

\item{[5]} S. Hojman, S. Chayet, D. N\'u\~nez, and M. Roque, J. Math.
Phys. {\bf 32}, 1491 (1991).

\item{[6]} In Weyl coordinates, these equations take the form
$4 \gamma_\rho= \rho f^{-2} (f_\rho^2 - f_z^2) - \rho^{-1}f^2
(\omega_\rho^2
-\omega_z^2)$ and $2\gamma_z  = \rho f^{-2}f_\rho f_z -
\rho^{-1} f^2\omega_\rho \omega_z$, with $f=\exp(2\psi)$.

\item{[7]} F.J. Ernst, Phys. Rev. {\bf 167}, 1175 (1968).

\item{[8]} Y. Choquet--Bruhat, C. DeWitt--Morette, and M.
Dillard--Bleick,
{\it Analysis, Manifolds and Physics}, (North--Holland, Amsterdam,
1982).

\item{[9]} G. Fodor, C. Hoenselares, and Z. Perj\'es, J. Math. Phys.
{\bf 30},
2252 (1989).

\item{[10]} See, for instance, {\it Solutions of Einstein's
equations:

Techniques
and Results}, edited by C. Hoenselaers and W. Dietz (Springer,
Berlin, 1984).
For an introductory review see H. Quevedo, Fortschr. Phys. {\bf 38},
733
(1990).

\item{[11]} P.A. Dirac, Proc. Roy. Soc. A,  {\bf 133}, 60 (1931);
Phys. Rev.
{\bf 74}, 817 (1948).

\item{[12]} G. t'Hooft, Nucl. Phys. B, {\bf 79}, 276 (1974); A. M.
Polyakov,
JETP Letters, {\bf 20}, 194 (1974).

\item{[13]} W. A. Hiscock, Phys. Rev. Lett. {\bf 50}, 1734 (1983).

\item{[14]} R. Geroch, J. Math. Phys. {\bf 11}, 1955 (1970); {\bf
11}, 2580
(1970); R.O. Hansen, {\it ibid.} {\bf 15}, 46 (1974). For other
equivalent
definitions of multipole moments see:  K. S. Thorne, Rev. Mod. Phys.
{\bf 52}, 299 (1980); R. Beig and W. Simon, Comm. Math. Phys. {\bf
78}, 75
(1980); Proc. Roy. Soc. London, {\bf 376A}, 333 (1981); Acta Phys.
Austriaca
{\bf 53}, 249 (1981).
For an introductory review see H. Quevedo, Fortschr. Phys. {\bf 38},
733
(1990).

\item{[15]} It can be shown that a stationary axisymmetric solution
is
asymptotically flat (see Ref. [14]) if for $\rho=0$ and
$z\rightarrow\infty$
the metric functions behave like $f\rightarrow 1 + O(z^{-1})$ and
$\Omega \rightarrow O(z^{-1})$.

\item{[16]} J. Chazy, Bull. Soc. Math. France, {\bf 52}, 17 (1924);
H. E. J. Curzon, Proc. Math. Soc. London, {\bf 23}, 477 (1924).

\item{[17]} This is a special case of a more general solution
presented in: H. Quevedo, Phys. Rev. D {\bf 39}, 2904 (1989).

\item{[18]} H. Thirring and J. Lense, Phys. Z. {\bf 19}, 156 (1918);
see also
B. Mashhoon, F. W. Hehl, and  D. S. Theiss, Gen. Rel. Grav. {\bf 16},
711
(1984).

\bye